\documentclass[fleqn,twoside,twocolumn,nofootinbib,showkeys]{revtex4} 
\usepackage[sec,nocpr]{ujp} 

\begin{document}
\title[Diffractive Physics at the LHC]
{Diffractive Physics at the LHC}%
\author{Maciej Trzebi\'nski}
\affiliation{Institute of Nuclear Physics Polish Academy of Sciences}
\address{152 Radzikowskiego Str., 31-342 Cracow, Poland}
\email{maciej.trzebinski@ifj.edu.pl}

\udk{??} \razd{\secix}


\setcounter{page}{1}%

\begin{abstract}
Diffractive processes possible to be measured at the LHC are listed and briefly discussed. This includes soft (elastic scattering, exclusive meson pair production, diffractive bremsstrahlung) and hard (single and double Pomeron exchange jets, $\gamma$+jet, $W/Z$, jet-gap-jet, exclusive jets) processes as well as Beyond Stanrad Model phenomena (anomalous gauge couplings, magnetic monopoles).
\end{abstract}

\keywords{LHC, AFP, ALFA, TOTEM, Pomeron, diffraction, exclusive processes, beyond standard model}

\maketitle

\section{Introduction}
About a half of collisions at the LHC are of diffractive nature. In such events a rapidity gap\footnote{A space in rapidity devoid of particles.} between the centrally produced system and scattered protons is present. Due to an exchange of colourless object -- photon (electromagnetic) or Pomeron (strong interaction) -- one or both outgoing protons may stay intact.

Studies of diffractive events are an important part of the physics programme of the LHC experiments. Diffractive production could be recognized by a search for a rapidity gap in the forward direction or by measurement of scattered protons. The first method is historically a standard one for diffractive pattern recognition. It uses the usual detector infrastructure: \textit{i.e.} tracker and forward calorimeters. Unfortunately, rapidity gap may be destroyed by \textit{e.g.} particles coming from pile-up -- parallel, independent collisions happening in the same bunch crossing. In addition, gap may be outside the acceptance of detector. In the second method protons are directly measured. This solves the problems of gap recognition in the very forward region and a presence of pile-up. However, since protons are scattered at small angles (few hundreds micro-radians), additional devices called ``forward detectors'' are needed to be installed.

At the LHC the so-called Roman pot technology is used. In ATLAS \cite{ATLAS} two systems of such detectors were installed: ALFA \cite{ALFA1, ALFA2} and AFP \cite{AFP}. At the LHC interaction point 5, Roman pots are used by CMS \cite{CMS} and TOTEM \cite{TOTEM1, TOTEM2} groups. Since protons are scattered at small angles, there are several LHC elements (\textit{i.e.} magnets and collimators) between them and the IP which influence their trajectory. Settings of these elements, commonly called machine optics, determines the acceptance of forward detectors. The detailed description of the properties of optics sets used at the LHC can be found in \cite{Trzebinski_optics}.

In both experiments, a large community works on both phenomenological and experimental aspects of diffraction. In this paper, the diffractive processes possible to be measured will be briefly described.

\section{Soft Diffraction}
Collisions at hadron accelerators are dominated by soft processes. Absence of a hard scale in these events prevents one from using perturbation theory. Instead, in order to calculate the properties of the produced particles such as energy or angular distributions, one has to use approximative methods.

The elastic scattering process has the simplest signature that can be imagined: two protons exchange their momentum and are scattered at small angles. At the LHC, measurement of protons scattered elastically requires a special settings commonly named the high-$\beta^*$ optics. Properties of elastic scattering were measured by both ATLAS and TOTEM Collaborations for center-of-mass energy of 7 \cite{ATLAS_elastic_7TeV, TOTEM_elastic_7TeV}, 8 \cite{ATLAS_elastic_8TeV, TOTEM_elastic_8TeV} and 13 TeV \cite{TOTEM_elastic_13TeV}.

Another soft process is a diffractive bremsstrahlung. It is typically of electromagnetic nature. However, high energy photons can be radiated in the elastic proton-proton scattering as postulated in \cite{brem1}. This idea was further extended in \cite{brem2} by introducing the proton form-factor into the calculations and by considering other mechanisms such as a virtual photon re-scattering. Feasibility studies presented in \cite{brem3} suggest that such measurement should be possible at the LHC. The requirements are high-$\beta^*$ optics, proton measurement in ALFA/TOTEM and photon measurement in Zero Degree Calorimeter.

Last of processes described in this Section is the exclusive meson pair production, a $2 \to 4$ process in which two colliding protons result in two charged mesons and two scattered protons present in the final state. In the non-resonant pion pair production (also called continuum) a Pomeron is ``emitted'' from each proton resulting in a four particles present in final state: scattered protons and (central) pions \cite{LS1}. The object exchanged in the $t$-channel is an off-shell pion. Exclusive pions can also be produced via resonances, \textit{e.g.} $f_0$ \cite{LS2}. Although the dominant diagram of the exclusive pion pair continuum production is a Pomeron-induced one, the production of a photon-induced continuum is also possible. On top of that, a resonant $\rho^0$ photo-production process may occur \cite{LS3}.

Recently, the models of the elastic scattering, exclusive meson production and diffractive bremsstrahlung were added to the GenEx Monte Carlo generator \cite{GenEx1, GenEx2, GenEx3}.

\section{Hard Diffraction}
Hard diffractive events can be divided into a single diffractive and double Pomeron exchange classes. In the first case, one proton stays intact whereas the other one dissociates. In the second case, both interacting protons ``survive''. In addition, the sub-case of the exclusive production can considered -- a processes in which all final state particles can be measured by ATLAS and CMS/TOTEM detectors.

Depending on the momentum lost during the interaction, the emitting proton may remain intact and be detected by a forward proton detector. However, it may happen that soft interactions between the protons or the proton and the final state particles can destroy the diffractive signature. Such effect is called the gap survival probability. For the LHC energies gap survival is estimated to be of about 0.03 -- 0.1 depending on the process \cite{Khoze_gap}.

From all hard events, diffractive jets have the highest cross-section\footnote{Depends on the jet transverse momentum.}. By studying a single diffractive and double Pomeron exchange jet production, a Pomeron universality between $ep$ and $pp$ colliders can be probed. As was discussed in \cite{Royon_jets}, tagging of diffractive protons will allow the QCD evolution of gluon and quark densities in the Pomeron to be tested and compared to the ones extracted from the HERA measurements. Another interesting measurement is the estimation of the gap survival probability. A good experimental precision will allow for comparison to theoretical predictions and differential measurements of the dependence of the survival factor on \textit{e.g.} mass of the central system. 

An interesting class of jet events is one with a rapidity gap is present between jets -- so-called jet-gap-jet production. In such events, an object exchanged in the $t$-channel is a colour singlet and carries a large momentum transfer. When the gap size is sufficiently large, the perturbative QCD description of jet-gap-jet events is usually performed in terms of the Balitsky-Fadin-Kuraev-Lipatov (BFKL) model \cite{BFKL1, BFKL2, BFKL3}. The jet-gap-jet topology can be produced also in the single diffractive and double Pomeron exchange processes. Properties of such events were never measured -- the determination of the cross-section should enable the tests of the BFKL model \cite{BFKL4}.

Jets produced in processes described above are typically of gluonic nature. In order to study the quark composition of a Pomeron, diffractive photon+jet productions should be considered. In such cases one Pomeron emits a gluon, whereas the other one delivers a quark. A measurement of photon+jet production in DPE mode can be used to test the Pomeron universality between HERA and LHC. Moreover, HERA was not sensitive to the difference between the quark components in the Pomeron. It means that the fits assumed equal amount of light quarks, $u = d = s = \bar{u} = \bar{d} = \bar{s}$. The LHC data should allow more precise measurements \cite{gamma_jet}.

Another interesting process is diffractive production of $W$ and $Z$ bosons. Similarly to $\gamma$+jet, it is sensitive to the quark component since many of the observed production modes can originate from a quark fusion. As was discussed in \cite{diffractive_WZ}, by measuring the ratio of $W$ to $Z$ cross-section the d/u and s/u quark density values in the Pomeron can be estimated. In addition, a study of the DPE $W$ asymmetry can be performed \cite{diffractive_WZ}. Such measurement can be used to validate theoretical models.

Seasibility studies of all measurements described above in this section are describled in Ref. \cite{LHC_forward_physics}.

Diffractive jets can be produced in the exclusive mode. Usually, it is assumed that one gluon is hard, whereas the other one is soft \cite{Khoze_exc_jj, Albrow_exc_jj}. The role of the soft gluon is to provide the colour screening in order to keep the net colour exchange between protons equal to zero. The exclusivity of the event is assured via the Sudakov form factor \cite{Sudakov}, which prohibits an additional radiation of gluons in higher orders of perturbative
QCD. In \cite{EXC_DT}, a discussion about feasibility of such measurement for case of the ATLAS detector and both tagged protons is held. A semi-exclusive measurement, when one of protons is tagged, is discussed in \cite{EXC_ST1, EXC_ST2}.

\section{Anomalous Couplings and Beyond Standard Model Physics}
Presence of intact proton can be used to search for a new phenomena. Beyond Standard Model processes are usually expected to be on a high mass, which makes them visible in forward detectors. 

One example BSM physics are anomalous couplings: $\gamma \gamma W W$, $\gamma \gamma Z Z$, $\gamma \gamma \gamma \gamma$ or $WW\gamma$. As was shown in \cite{anomalous_couplings1, anomalous_couplings2}, the possibility of forward proton tagging provides much cleaner experimental environment which improves the discovery potential. Authors expect that with 30 - 300 fb$^{-1}$ of data collected with the ATLAS detector with information about scattered protons tagged in AFP should result in a gain in the sensitivity of about two orders of magnitude over a standard ATLAS analysis.

Finally, a presence of protons with high energy loss and a lack of energy registered in the central detector might be a sign of a new physics, for example a magnetic monopoles \cite{LHC_forward_physics}.

\section{Conclusions}
Large Hadron Collider gives a possibility to study the properties of diffractive physics in a new kinematic domain. Diffractive events can be identified in all major LHC experiments using the rapidity gap recognition method. In addition, as ATLAS and CMS/TOTEM are equipped with the set of forward detectors, it is possible to use the proton tagging technique.

In this paper a brief summary of diffractive processes measurable at the LHC was done. Using special settings of the LHC -- high-$\beta^*$ optics -- processes of elastic scattering, exclusive meson pair production and diffractive bremsstrahlung can be studied. Hard diffractive events, due to smaller cross-sections, should be measured with the standard LHC optics. Determination of properties of diffractive di-jet, photon+jets and $W/Z$ boson production processes should lead \textit{i.a.} to determination of gap survival probability and Pomeron structure. Studies of diffractive jet-gap-jet events should bring more insight into description of Pomeron, \textit{i.a.} to verify predictions of the BFKL model. On top of that, the measurement of jet production in the exclusive (double proton tag) and semi-exclusive (single tag) mode can be performed. Finally, additional information about scattered proton may improve the searches for a New Physics -- phenomena as anomalous gauge couplings or magnetic monopoles.

\vskip3mm \textit{Development of the GenEx Monte Carlo generator was partially founded by the Polish National Science Center grant: UMO-2015/17/D/ST2/03530.}

\end{document}